\documentclass[a4paper,fleqn,usenatbib]{mnras}

\usepackage{newtxtext,newtxmath}

\usepackage[T1]{fontenc}
\usepackage{ae,aecompl}

\usepackage{graphicx}   
\usepackage{amsmath}    
\usepackage{amssymb}    
\usepackage{amsfonts}
\usepackage{bm,color}

\newcommand{\der}[2]{\frac{\partial #1}{\partial #2}}
\newcommand{\dder}[3]{\frac{\partial^2 #1}{\partial #2\partial #3}}

\newcommand{\w}[1]{\bm{#1}}
\newcommand{\vv}[1]{\vec{\w{#1}}}

\newcommand{\be}{\begin{equation}}
\newcommand{\ee}{\end{equation}}
\newcommand{\bea}{\begin{eqnarray}}
\newcommand{\eea}{\end{eqnarray}}

\newcommand{\Liec}[1]{{\mathcal{L}}_{\vv{#1}}\,}

\title[Supermassive Neutron Stars in Axion $F(R)$ Gravity]{Supermassive Neutron Stars in Axion $F(R)$ Gravity}
\author[Astashenok \& Odintsov]{Artyom V. Astashenok$^{1}$, Sergey D. Odintsov$^{2,3,4,5}$\\
\small $^{1}$Immanuel Kant Baltic Federal University\\
\small Department of Physics, Technology and IT\\
\small 236041 Kaliningrad, Russia, Nevskogo str.14\\
\small $^{2}$Institut de Ci\'{e}ncies de l'Espai, ICE/CSIC-IEEC, Campus UAB, Carrer de Can Magrans s/n, 08193 Bellaterra (Barcelona), Spain \\
\small $^{3}$Instituci\'{o} Catalana de Recerca i Estudis
Avan\c{c}ats (ICREA), Barcelona, Spain\\
\small $^{4}$Int. Lab. Theor. Cosmology, TUSUR, 634050 Tomsk, Russia\\
\small $^{5}$Institute of  Physics, Kazan Federal University,
Kazan 420008, Russia}

\begin{document}
\label{firstpage}
\pagerange{\pageref{firstpage}--\pageref{lastpage}} \maketitle

\maketitle

\begin{abstract}
We investigated  realistic neutron stars in axion $R^{2}$ gravity.
The coupling between curvature and axion field $\phi$ is assumed
in the simple form $\sim R^2\phi$. For the axion mass in the range
$m_{a}\sim 10^{-11}-10^{-10}$ eV the solitonic core within neutron
star and corresponding halo with size $\sim 100$ km can exist.
Therefore the effective contribution of $R^2$ term grows inside
the star and it  leads to change of star parameters (namely, mass
and radius). We obtained the increase of star  mass independent
from central density for wide range of masses. Therefore, maximal
possible mass for given equation of state grows. At the same time,
the star radius increases not so considerably in comparison with
GR. {Hence, our model may predict possible existence of
supermassive compact stars with masses $M\sim 2.2-2.3M_\odot$ and
radii $R_{s}\sim 11$ km for realistic equation of state (we
considered APR equation of state). In General Relativity one can
obtain neutron stars with such characteristics only for
unrealistic, extremely stiff equations of state.} Note that this
increase of mass occurs due to change of solution for scalar
curvature outside the star. In GR curvature drops to zero on star
surface where $\rho=p=0$. In the model under consideration the
scalar curvature dumps more slowly in comparison with vacuum $R^2$
gravity due to axion ``galo'' around the star.
\end{abstract}


\section{Introduction}

There are still unresolved fundamental puzzles in modern cosmology
and relativistic astrophysics. One of them is so-called dark
energy which governs  the observed accelerated universe expansion
(\citealp{Riess-1}; \citealp{Perlmutter}; \citealp{Riess-2}).
According to the well-known $\Lambda $CDM model, dark energy is
simply Cosmological Constant and its density is 72\% of the global
energy budget of the universe. The remaining 28\%, clustered in
galaxies and clusters of galaxies, consist of baryons (only 4\%)
and cold dark matter (CDM) the nature of which is unclear.
Alternative approach to description of late-time cosmological
dynamics of universe is proposed by  theories of modified gravity
(\citealp{Capozziello1}; \citealp{Odintsov1}; \citealp{Turner}).

Furthermore, the unified description of observed late-time
acceleration and early universe expansion  is also possible in the
context of $f(R)$ theory (\citealp{Nojiri}). Matter and radiation
dominance eras can be also described in frames of this approach
(for review, see refs. \citealp{Nojiri-5}; \citealp{Capoz};
\citealp{Olmo-2}; \citealp{Cruz}; \citealp{Nojiri-4}).

Another problem of current cosmology and high energy physics is
dark matter. There are many evidences in favor for particle nature
of dark matter (for example, see data about collision of galaxies
in the Bullet Cluster and {cluster MACSJ0025}
(\citealp{Markevitch}; \citealp{Clowe}; \citealp{Robertson};
\citealp{Bradac})). As the candidates on the role of dark matter
the weakly interacted massive particles (WIMPs) have been
considered usually. Many approaches to direct observation of such
particles were proposed but so far the direct experiments for
detection of such particles have not been successful
({\color{blue} see} \citealp{Ahmed}; \citealp{Davis};
\citealp{Davis2}; \citealp{Roszkowski}; \citealp{Schumann}).

Another possibility is that dark matter is nothing else than
axions (\citealp{Sakharov}; \citealp{Sakharov-2};
\citealp{Sakharov-3}; \citealp{Marsh}; \citealp{Marsh-2};
\citealp{Oikonomou}; \citealp{Cicoli}; \citealp{Fukunaga};
\citealp{Caputo}). Some recent experiments indicate in favor for
its existence (\citealp{ADMX}; \citealp{ABRACADABRA};
\citealp{Quellet}; \citealp{Safdi}; \citealp{Avignone};
\citealp{Caputo-2}; \citealp{Caputo-3}; \citealp{AX-1};
\citealp{Rozner}). Mass of axions can be very low (theoretical
estimations give value in the wide range $\sim 10^{-12}-10^{-3}$
eV). The possibility of axions detection is based on axion-photon
interaction in the presence of magnetic fields (\citealp{AX-3};
\citealp{AX-4}; \citealp{AX-5}).

Recently, axion F(R) gravity was discussed in ref.
\citealp{Oikonomou-19} in which the unification of dark energy with axion dark matter (and
eventually, with inflation) was proposed. For the description of
axion  in this model misalignment model was used (\citealp{AX-2}).
According to misagliment model the primordial U(1) Peccei-Quinn
symmetry is broken during inflation. For $f(R)$ it was chosen
simple $R^2$ model and non-minimal coupling with axion field  in
the form $h(\phi)R^{\gamma}$.

It is interesting to address the question about possible
manifestations of modified gravity with various scalar fields
(including axions) on astrophysical level. First of all, one
should pay attention to relativistic stars for which the energy
density in the center is around $10^{15}-10^{16}$ g/cm$^{3}$ and
therefore gravitational field is extremely high. In principle, for
such strong gravity regime the possible deviations from General
Relativity can be visible somehow. Unfortunately we have no well
established data about mass-radius diagram for neutron stars from
astronomical observations. Mass, radius and other parameters of
relativistic stars also depend from the equation of state chosen
for dense matter. There is now clear understanding how nuclear
matter behaves at such extremely densities and tens of EoS were
proposed over the years (for some introduction, see ref.
\citealp{Rezzolla}).

This paper is devoted to the study   of neutron stars in frames of
  $R^2$ gravity  coupled with axion field
$\phi$ in the form $\beta R^2\phi$ where $\beta$ is some constant.
It is interesting to note that Compton wavelength $\lambda_{a}$
for particle with mass $m_{a}\sim 10^{-11}-10^{-10}$ eV is $\sim
10-10^2$ km. This scale is comparable with characteristic size of
neutron stars. Therefore one can consider the possibility of
existence of solitonic core containing dark matter in the center
of the star. The contribution to energy density from such core is
negligible  itself and therefore couldn't influence on the
parameters of star. However, the assumption of coupling $\sim
R^2\phi$ can lead to non-trivial deviations from General
Relativity.

$F(R)$ gravity was considered as viable alternative to GR for
description of stellar structure in some papers. The question of
hydrostatic equilibrium was studied in ref. \citealp{Capozz2011}.
Dynamics and collapse of collisionless self-gravitating systems in
$R^2$ gravity was firstly investigated by \citealp{Capozz2012}. It
is interesting to note also iterative procedure for the solution
of modified Lane-Emden equation considered in \citealp{Capozz-2}.
For neutron stars initially the perturbative approach was used.
The scalar curvature $R$ is defined by Einstein equations at
zeroth order on the small parameter, i.e. $R \sim T$, where $T$ is
the trace of energy-momentum tensor. This approach is applied to
construction of neutron star models in $f(R)=R+\alpha R^2+\beta
R^3$ and $f(R)=R+\alpha R^2(1+\gamma \ln R)$ gravity also in
\citealp{Arapoglu2011}, \citealp{Alavirad2013};
\citealp{Astashenok2013}.

In modified $f(R)$ gravity model with cubic and quadratic terms,
it is possible to obtain neutron stars with $M\sim 2M_{\odot}$ for
simple hyperon equations of state (EoS) although the soft hyperon
equation of state is usually treated as non-realistic in the
standard General Relativity (\citealp{Astashenok2014}). The possible
signatures of modified gravity in neutron star astrophysics also
can include existence of neutron stars with extremely high
magnetic fields (\citealp{Cheoun2013}; \citealp{Astashenok2015}). Interesting
results were obtained for $f(R)=R^{1+\epsilon}$ gravity in
\citealp{Capozz-1}. Detailed analysis of neutron stars structure in
$R^2$ gravity was given by authors in ref. \citealp{Astashenok2017}.
It is shown that so-called gravitational sphere with nonzero
curvature appears around the star. Recently the mass-radius
relation in both metric and torsional $R^2$ gravity were
investigated in \citealp{Capozz-3}. For recent review of compact star
models in modified theories of gravity see \citealp{Olmo} and
references therein.

The main problem of simple $R^2$ gravity is that possible
observable consequences appear only if the contribution of
$R^2$-term  is sufficiently large. The motivation for the
consideration of non-minimal coupling in the form $\sim R^2\phi$
is that axion field can strengthen the contribution of $R^2$-term
only inside star due to solitonic core. Outside the star the
scalar field and scalar curvature quickly drop in fact to zero
value (of course, in comparison with corresponding values inside
the  star).

In the next section we start from the equations for stellar
configurations in $f(R)$ gravity in quasi-isotropic coordinates.
For their derivation the so-called $3+1$ formalism is used. as a
result we get the  system of equations of elliptical type for
unknown metric functions and scalar curvature $R$. The account  of
scalar field requires one more equation. The results of the
calculation including mass profile, mass-radius diagram, scalar
curvature and axion field are given in the section III. For
neutron star matter the well-known APR EoS is used. Our
calculations show that qualitative behavior of solutions doesn't
depend from chosen EoS.

\section{3+1 formalism in f(R) gravity}

We start from the well-known Einstein equations in frames of
General Relativity
\begin{equation}
R_{\mu\nu}-\frac{1}{2}g_{\mu\nu}R={8\pi} T_{\mu\nu}.
\end{equation}
Here $R_{\mu\nu}$ is the Ricci tensor associated with the
Levi-Civita connection $\nabla$ in 4-dimensional spacetime,
$R=g^{\mu\nu}R_{\mu\nu}$ is the scalar curvature and $T_{\mu\nu}$
is the energy-momentum tensor of matter. The system of units in
which $G=c=1$ is used.

Firstly we consider simple $f(R)$ gravity with the action
\begin{equation}\label{EinFR}
S=\frac{1}{2}\int f(R) \sqrt{-g} d^{4} x
\end{equation}
from which it follows that Einstein equations are
\begin{equation}
f_{R}(R) R_{\mu\nu}-\frac{f(R)}{2} \, g_{\mu\nu} -\left(
\nabla_{\mu} \nabla_{\nu}- g_{\mu\nu}\Box\right) f_{R}(R))=8\pi
T_{\nu\mu}.
\end{equation}
Here $\Box=\nabla^{\mu}\nabla_{\mu}$ is covariant D'Alamber
operator and $f_{R}(R)\equiv df/dR$. Then we drop arguments of
$f(R)$.

One can rewrite (\ref{EinFR}) in equivalent form
\begin{equation}\label{EinFR2}
f_{R}R_{\mu\nu}-\frac{1}{2}(f_{R}R-f)g_{\mu\nu}-\left(\frac{1}{2}\Box+\nabla_{\mu}
\nabla_{\nu}\right)f_{R}=8\pi
\left(T_{\mu\nu}-\frac{1}{2}g_{\mu\nu}T\right),
\end{equation}
where $T$ is the trace of energy-momentum tensor.

For the study of compact stars in general relativity the 3+1
formalism is often used (\citealp{Gourg}; \citealp{Alcubierre};
\citealp{Shapiro}; \citealp{Gourg-2}). One can adopt this method
for $f(R)$ gravity without any significant changes.

The key moment is foliation of spacetime by spacelike
hypersurfaces $\Sigma_{t}$. Parameter $t$ can be associated with
coordinate time. The next step is the definition of metric
$\gamma_{ij}$ induced by metric $g_{\mu\nu}$ on hypersurface
$\Sigma$. The components of induced metric can be given via the
components of unit timelike normal vector $\textbf{n}$ and metric
$g_{\mu\nu}$:
\begin{equation}
\gamma_{\alpha\beta}=g_{\alpha\beta}+n_{\alpha}n_{\beta}.
\end{equation}
We also define the components of orthogonal projector onto
hypersurface $\Sigma_{t}$ by raising of the first index:
\begin{equation}
\gamma^{\alpha}_{\cdot\beta}=\delta^{\alpha}_{\cdot\beta}+n^{\alpha}
n_{\beta}.
\end{equation}

We use metric of special form
\begin{equation}
ds^{2}=-N^{2}dt^{2}+\gamma_{ij}(dx^{i}+\xi^{i}dt)(dx_{j}+\xi_{j}dt),
\end{equation}
where $N$ is so-called lapse function and $\vv{\xi}$ is shift
vector. The 3-dimensional metric $\gamma_{ij}$ is the metric
induced on surface. Then projecting the Einstein equation
(\ref{EinFR2}) twice onto $\Sigma_t$, (ii) twice along to vector
$\vv{n}$ normal to $\Sigma_{t}$ and (iii) once on $\Sigma_t$ and
once along $\vv{n}$, one gets the following three equations:
\begin{equation} \label{PDE1}
f_{R}\left(\der{K_{ij} }{t} - \Liec{\xi} K_{ij}\right)
     +f_{R}\left(D_i D_j N - N\left\{
     {}^3 R_{ij} + K K_{ij} -2 K_{ik} K^k_{\ \, j}\right\}\right)=
\end{equation}
$$
=4\pi N \left[ (\sigma-\epsilon) \gamma_{ij} - 2 \sigma_{ij}
\right]-\frac{1}{2}(f_{R}R-f)N\gamma_{ij}-N\left(\frac{1}{2}\gamma_{ij}\Box
+D_{i}D_{j}\right)f_{R},
$$
\begin{equation}
f_{R}({}^3 R + K^2 - K_{ij} K^{ij}) = 16\pi E+f_{R}R-f+2D^{i}
D_{i} f_{R},
  \label{PDE2}
\end{equation}

\begin{equation}
f_{R}( D_j K^j_{\ \, i} - D_i K) = 8\pi  p_i
-n^{\mu}\nabla_{\mu}(D_{i}f_{R}), \label{PDE3}
\end{equation}

where $K_{ij}$ is tensor of extrinsic curvature and $K=K^{i}_{i}$.
One also introduces the components of Lie derivative of tensor and
$K_{ij}$ along the vector $\vv{\xi}$: \be \label{e:ein:Lie_beta_K}
    \Liec{\xi} K_{ij} = \xi^k \der{K_{ij}}{x^k}
     + K_{kj} \der{ \xi^k}{x^i} + K_{ik} \der{\xi^k}{x^j}.
\ee

The covariant derivatives $D_i$ can be expressed in terms of
partial derivatives with respect to the spatial coordinates
$(x^i)$ by means of the Christoffel symbols ${}^3 \Gamma^i_{\ \,
jk}$ of $\w{D}$ associated with $(x^i)$:
\be
   D_i D_j N = \dder{N}{x^i}{x^j} - {}^3 \Gamma^k_{\ \, ij}
     \der{N}{x^k} ,
\ee
\be
D_j K^j_{\ \, i} = \der{K^j_{\ \, i}}{x^j}
     + {}^3 \Gamma^j_{\ \, jk} K^k_{\ \, i}
    - {}^3 \Gamma^k_{\ \, ji} K^j_{\ \, k} ,
\ee
\be
   D_i K = \der{K}{x^i} .
\ee

For 3-dimensional Ricci tensor $^{3}R_{ij}$ and scalar curvature
$^{3}R$ we have following relations:

\be {}^3 R_{ij} = \der{\, {}^3 \Gamma^k_{\ \, ij}}{x^k} - \der{\,
{}^3 \Gamma^k_{\ \, ik}}{x^j}
     + {}^3 \Gamma^k_{\ \, ij} {}^3 \Gamma^l_{\ \, kl}
     - {}^3 \Gamma^l_{\ \, ik} {}^3 \Gamma^k_{\ \, lj}  \label{e:sym:3Ricci}
\ee
\be{}^3 R = \gamma^{ij} \, {}^3 R_{ij}.
   \label{e:sym:3Ricci_scal}
\ee

The quantities $\epsilon$, $S_{ij}$ and $p_{i}$ are energy
density, components of stress tensor and vector of energy flux
density correspondingly. These values can be obtained from
stress-energy tensor $T_{\mu\nu}$:
\begin{equation*}
\epsilon=n^{\mu} n^{\nu} T_{\mu\nu},
\end{equation*}
\begin{equation}
\sigma_{ij}=\gamma^{\mu}_{i}\gamma^{\nu}_{j} T_{\mu\nu},\quad
\sigma=\sigma^{i}_{i}.
\end{equation}
\begin{equation*}
p_{i}=-n^{\mu}\gamma^{\nu}_{i}T_{\mu\nu}.
\end{equation*}
Then we take the trace of first equation:
\begin{equation}
f_{R}D_{i}D^{i}N=Nf_{R}({}^{3}R+K^{2}-2K_{ik}K^{ik})+4\pi N
(S-3E)+
\end{equation}
$$
+\frac{3}{2}N(f-f_{R}R)-\frac{3}{2}N\Box f_{R}-ND_{i}D^{i}f_{R}.
$$
From equation (\ref{PDE2}) it follows that
$$
f_{R}({}^{3}R+K^{2})=f_{R}K_{ij}K^{ij}+16\pi
E+f_{R}R-f+2D_{i}D^{i}f_{R}
$$
and therefore one can rewrite the previous equation as
\begin{equation}\label{PDE4}
f_{R}D_{i}D^{i}N=Nf_{R}K_{ij}K^{ij}+4\pi
N(\epsilon+\sigma)-\frac{1}{2}N(f_{R}R-f)-
\end{equation}
$$
-\frac{3}{2}N\Box f_{R}+ND_{i}D^{i}f_{R}.
$$

Let us consider the non-rotating stellar configurations. In this
case all metric functions depend only from radial coordinate. For
convenience we used isotropic spatial coordinates with metric in
the form
\begin{equation}\label{MetrSt}
ds^{2}=-N^{2}(r)dt^{2}+A^{2}(r)(dr^{2}+r^{2}d\Omega^{2}).
\end{equation}
For this metric one can obtain that
$$
K=0,\quad K_{i j} K^{i j}=0.
$$
We also have the following expression for 3-dimensional scalar
curvature:
\begin{equation}
   {}^3 R = -\frac{4}{A^3}
   \left( \frac{d^{2}A}{dr^{2}} + \frac{2}{r} \frac{dA}{dr}
   - \frac{1}{2A} \left( \frac{dA}{dr} \right) ^2 \right) \nonumber
\end{equation}
$$
$$
Taking into account that
\begin{equation*}
\frac{1}{A}\frac{d^{2}A}{dr^{2}}=\frac{d^{2}\ln
A}{dr^{2}}+\left(\frac{d\ln A}{dr}\right)^{2},
\end{equation*}
one can rewrite the equation for $^{(3)}R$ in the following form
\begin{equation}
{}^3 R = -\frac{4}{A^2}\left(\triangle^{r}_{(3)}\ln
A+\frac{1}{2}\left(\frac{d\ln A}{dr}\right)^{2}\right).
\end{equation}
In this relation $\triangle^{r}_{(3)}$ is nothing else than radial
part of $3$-dimensional Laplace operator in Euclidean space.

The energy-momentum tensor in the case of spherical symmetry can
be presented in diagonal form $T_{\mu}^{nu}=\mbox{diag}(-\epsilon,
p, p, p)$ where $p$ is pressure of matter and therefore
$$
\sigma^{r}_{r}=\sigma^{\theta}_{\theta}=\sigma^{\phi}_{\phi}=p,
\quad \sigma=3p.
$$

The action of 3-dimensional covariant D'Alamber operator for any
scalar function $\Phi(r)$ depending only from radial coordinate is
reduced to
\begin{equation}
D^{i}D_{i}\Phi(r)=\frac{1}{A^{2}}\left(\triangle^{r}_{(3)}\Phi+\frac{d\ln
A}{dr}\frac{d\Phi}{dr}\right).
\end{equation}
For 4-dimensional d'Alambertian one obtains the simple relation:
\begin{equation}
\Box\Phi(r)=\frac{1}{A^{2}}\left(\triangle^{r}_{(3)}\Phi+\frac{d\ln
(NA)}{dr}\frac{d\Phi}{dr}\right).
\end{equation}
Finally Eq. (\ref{PDE4}) for our task can be presented in the
form:
\begin{equation}\label{EQ1}
f_{R}\triangle^{r}_{(3)}\nu+\frac{1}{2}\triangle^{r}_{(3)}f_{R}=4\pi
A^{2}(\epsilon+3p)-\frac{A^{2}}{2}(f_{R}R-f)-f_{R}\frac{d\eta}{dr}\frac{d\nu}{dr}-
\end{equation}
$$
-\frac{1}{2}\frac{d\eta}{dr}\frac{df_{R}}{dr}-\frac{d\nu}{dr}\frac{df_{R}}{dr}.
$$
Here $\eta=\ln(AN)$ and $\nu=\ln N$. Let us consider eq.
(\ref{PDE2}) for metric (\ref{MetrSt}). Using relations for
$^{3}R$ and covariant d'Alambertian one gets
\begin{equation}\label{EQ2}
2f_{R}\triangle^{r}_{(3)}\ln A+\triangle^{r}_{(3)}f_{R}=-8\pi
A^{2} \epsilon-\frac{A^2}{2}(f_{R}R-f)-f_{R}\left(\frac{d(\ln
A)}{dr}\right)^{2}-
\end{equation}
$$
-\frac{d(\ln A)}{dr}\frac{df_{R}}{dr}.
$$

Next, one considers the $\phi\phi$-component of (\ref{PDE1}). One
need to use the following representations for $D_{\phi\phi}N$ and
$^{3}R_{\phi\phi}$:
$$
D_{\phi}D_{\phi}N=r^{2}\sin^{2}\theta\left(\frac{1}{A}\frac{dA}{dr}+\frac{1}{r}\right)\frac{dN}{dr},
$$
$$
^{3}R_{\phi\phi}=-r^{2}\sin^{2}\theta\frac{1}{A}\left(\frac{d^{2}A}{dr^{2}}+\frac{3}{r}\frac{dA}{dr}\right).
$$
One obtains after simple calculations
\begin{equation}\label{EQ3}
f_{R}\triangle^{r}_{(4)}\ln
A+\frac{1}{2}\triangle^{r}_{(3)}f_{R}=4\pi
A^{2}(p-\epsilon)-\frac{A^{2}}{2}(f_{R}R-f)-
\end{equation}
$$
-f_{R}\left(\frac{d\ln A}{dr}\right)^{2}-f_{R}\frac{d\ln
(Ar)}{dr}\frac{d\nu}{dr}-\frac{1}{2}\frac{d\eta}{dr}\frac{df_{R}}{dr}-\frac{d\ln
(Ar)}{dr}\frac{df_{R}}{dr}.
$$

Adding (\ref{EQ1}) and (\ref{EQ3}) gives the following equation
for $\eta$:
\begin{equation}\label{etaEQ}
f_{R}\triangle^{r}_{(4)}\eta+\triangle^{r}_{(4)}f_{R}=16\pi
A^{2}p-A^{2}(f_{R}R-f)-f_{R}\left(\frac{d
\eta}{dr}\right)^{2}-2\frac{d \eta}{dr}\frac{d f_{R}}{dr}
\end{equation}
Then adding (\ref{EQ1}) to (\ref{EQ2}) and subtracting (\ref{EQ3})
we obtain:

\begin{equation}
f_{R}\triangle^{r}_{(2)}\eta+\triangle^{r}_{(2)}f_{R}=8\pi
A^{2}p-\frac{1}{2}A^{2}(f_{R}R-f)-f_{R}\left(\frac{d
\nu}{dr}\right)^{2}-\frac{d \nu}{dr}\frac{d f_{R}}{dr}
\end{equation}

In General Relativity scalar curvature $R$ is simply $-8\pi T$ and
therefore it approaches to zero  on the surface of star where
$\epsilon=p=0$. For $f(R)$ gravity one needs additional equation
for scalar curvature. This equation can be obtained from trace of
Einstein equations and for our case of radial dependence of scalar
curvature it takes the form:
\begin{equation}\label{CurvEQ}
\triangle^{r}_{(3)}f_{R}=\frac{8\pi}{3}
A^{2}(3p-\epsilon)-\frac{A^{2}}{3}(f_{R}R-2f)-\frac{d
\eta}{dr}\frac{d f_{R}}{dr}.
\end{equation}

For the case of function $f_{R}=F(R,\phi)$ depending also from
scalar field $\phi$ these equations are valid. For radial
derivatives of function $F(R,\phi)$ one should remember that
\begin{equation*}
\frac{dF}{dr}=F_{R}\frac{dR}{dr}+F_{\phi}\frac{d\phi}{dr},
\end{equation*}
\begin{equation*}
\frac{d^{2}F}{dr^2}=F_{R}\frac{d^{2}R}{dr^2}+F_{RR}\left(\frac{dR}{dr}\right)^{2}+F_{\phi}\frac{d^{2}\phi}{dr^2}+F_{\phi\phi}\left(\frac{d\phi}{dr}\right)^{2}+
2F_{R\phi}\frac{dR}{dr}\frac{d\phi}{dr}.
\end{equation*}

Assuming the action for axion field in the following form
\begin{equation}
S_{\phi}=\int
d^{4}x\sqrt{-g}\left(-\frac{1}{2}\partial^{\mu}\phi\partial_{\mu}\phi-V(\phi)\right).
\end{equation}
one can obtain the equation for scalar field $\phi=\phi(r)$:
\begin{equation}\label{scalEQ}
\triangle^{r}_{(3)}\phi=A^{2}\frac{dV}{d\phi}-\frac{A^{2}}{8\pi}\frac{df}{d\phi}-\frac{d\phi}{dr}\frac{d\eta}{dr}.
\end{equation}

The system of equations (\ref{EQ1}), (\ref{etaEQ}),
(\ref{CurvEQ}), (\ref{scalEQ}) should be supplemented by a set of
boundary conditions for $\eta$, $\nu$, $R$ and $\phi$. Those are
provided by the asymptotic flatness assumption. On spatial
infinity the metric tensor tends towards Minkowski metric and
therefore
$$
\nu\rightarrow 0, \quad \eta\rightarrow 0, \quad R\rightarrow
0\quad \mbox{for}\quad r\rightarrow\infty.
$$
For scalar field we also assume that $\phi\rightarrow 0$ when
$r\rightarrow +\infty$ because the density of dark matter in the
space ($\sim 10^{-29}$ g/cm$^3$) is extremely low in comparison
with densities inside relativistic stars.

Asymptotical behavior of $A(r)$ at $r\rightarrow\infty$ defines
gravitational mass $M$ of star for distant observer. In General
Relativity the solution of Einstein equations outside the star has
the form:
\begin{equation}
A(r)=\left(1+\frac{M}{2r}\right)^{2}, \quad
N(r)=\left(1-\frac{M}{2r}\right)\left(1+\frac{M}{2r}\right)^{-1}.
\end{equation}

Therefore, the gravitational mass of star can be found  as an
asymptotical limit
\begin{equation*}
M=2\lim_{r\rightarrow\infty}r(\sqrt{A}-1).
\end{equation*}
One should also account that physical radial coordinate
$\tilde{r}$ is
\begin{equation*}
\tilde{r}=Ar.
\end{equation*}
Note that in the following the symbol ``r'' on figures means
physical distance. Tildes are omitted for simplicity.

\begin{figure}
\includegraphics[scale=0.32]{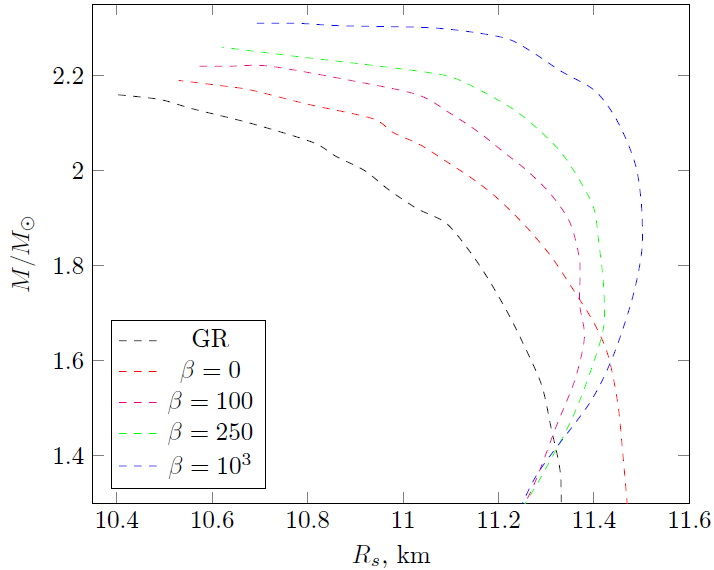}\\
\includegraphics[scale=0.32]{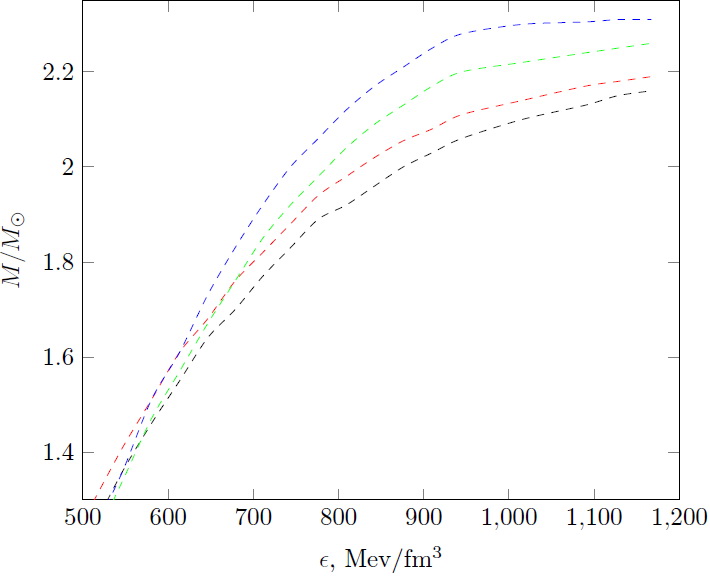}
\caption{Mass-radius diagram for $M>1.3M_{\odot}$ and dependence
mass from central energy density (in Mev/fm$^3$) for neutron stars
in General Relativity and model (\ref{our}) for various values of
$\beta$ and $\alpha=0.25$.}
\end{figure}

\begin{figure}
\includegraphics[scale=0.32]{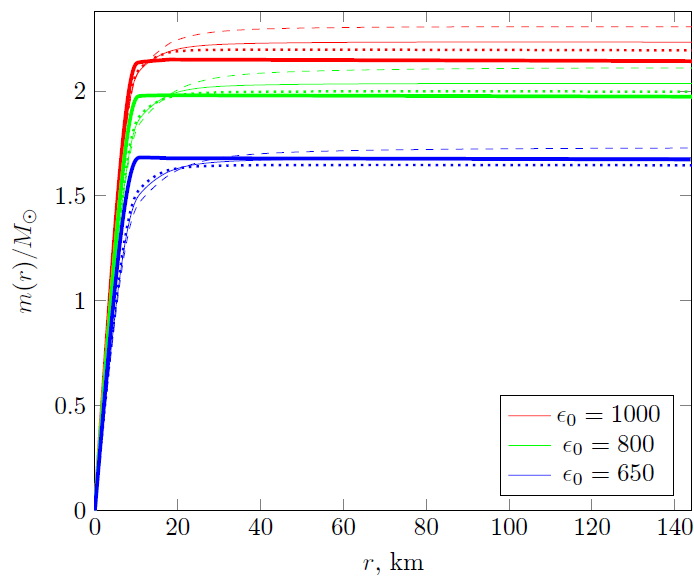}
\caption{The radial profile neutron star mass for case
$\alpha=0.25$ at $\epsilon_{0}=650$, $800$ and $1000$ Mev/fm$^3$
for $\beta=0$ (thick lines), $\beta=100$ (dotted lines),
$\beta=250$ (solid lines), $\beta=1000$ (dashed lines).}
\end{figure}

\begin{figure}
\includegraphics[scale=0.32]{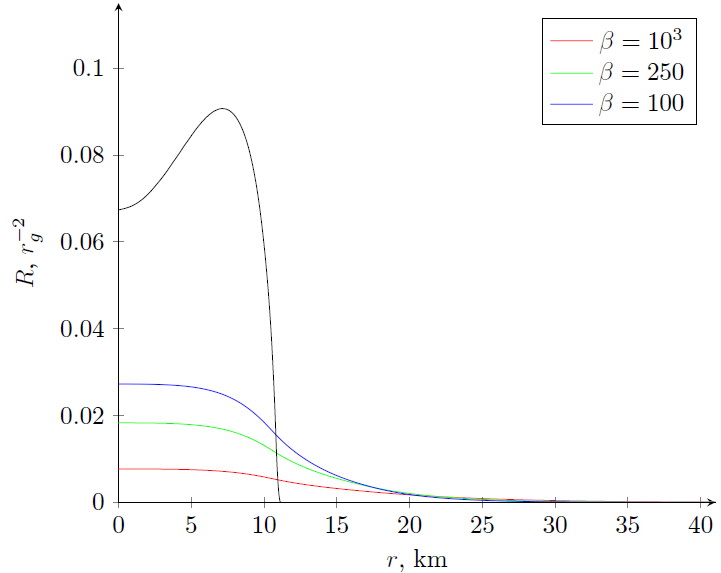}\\
\includegraphics[scale=0.32]{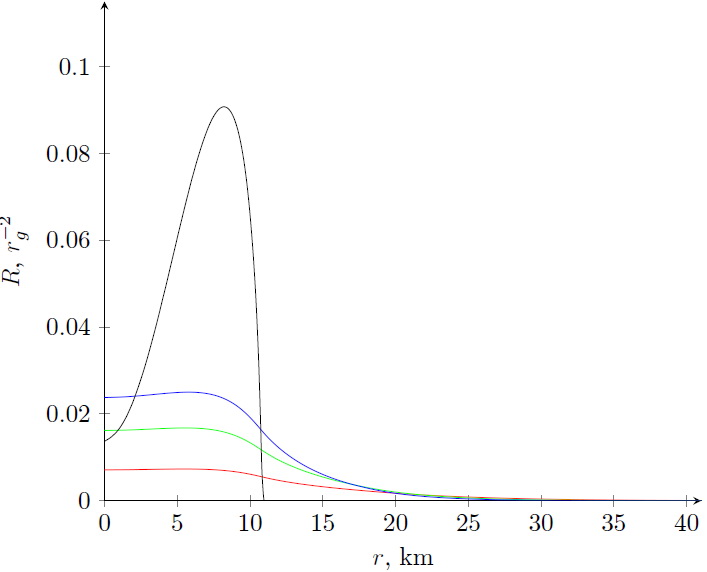}\\
\includegraphics[scale=0.32]{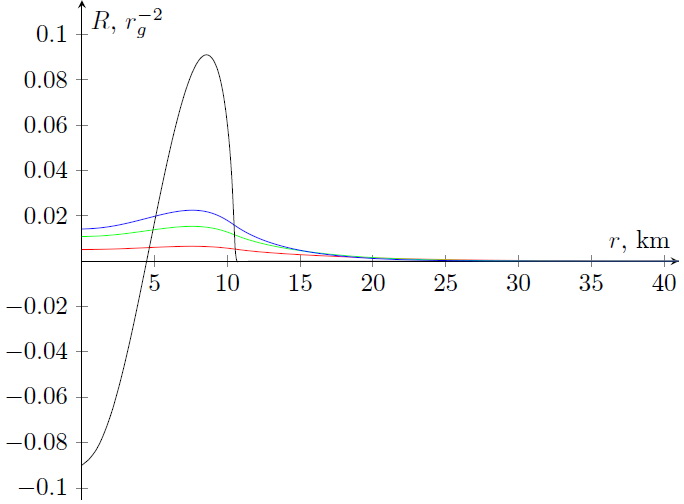}
\caption{The radial profile of curvature $R$ (in units of
$r_{g}^{-2}$) for case $\alpha=0.25$ at $\epsilon_{0}=650$ (upper
panel), $800$ (middle panel) and $1000$ Mev/fm$^3$ (down panel)
correspondingly. Black lines correspond to $R^2$ gravity without
axion field. {Note that for the case of $\epsilon_{0}=1000$
Mev/fm$^3$ curvature goes strongly negative in a range $r<4$ km.
This effect is only artefact of the choice of equation of state.
In simple $R^2$ gravity for our interval of $\alpha$ the solution
for scalar curvature $R$ doesn't significantly differ from the
solution in GR where scalar curvature is simply $8\pi(\rho-3p)$.
For APR EoS for large densities $3p>\rho$ and therefore $R<0$.}}
\end{figure}

\begin{figure}
\includegraphics[scale=0.32]{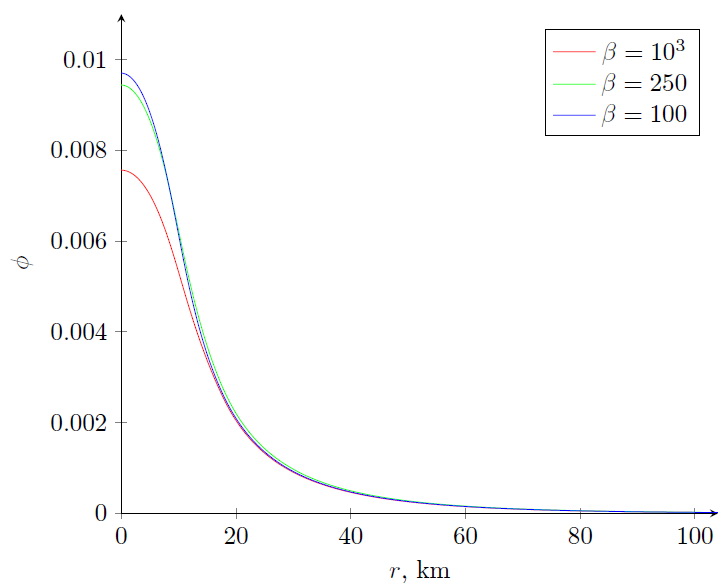}\\
\includegraphics[scale=0.32]{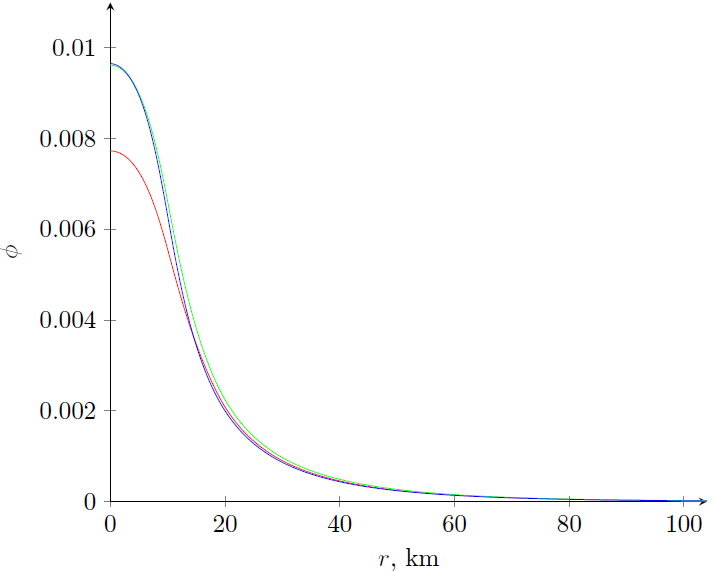}\\
\includegraphics[scale=0.32]{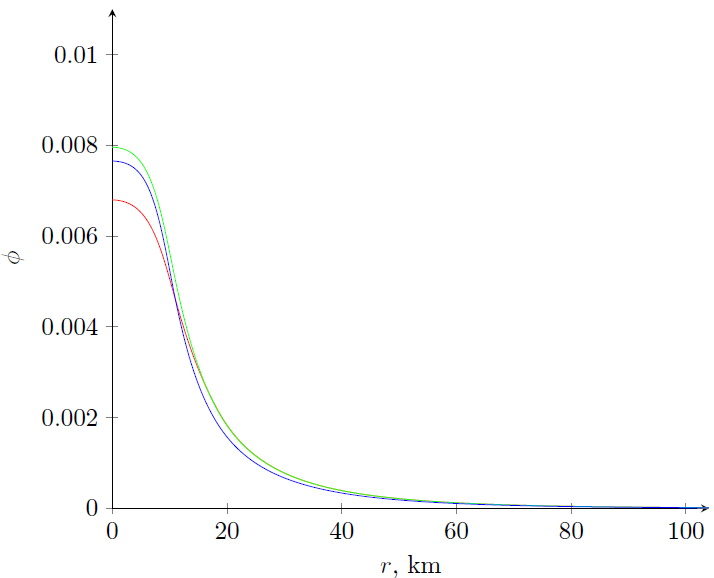}
\caption{The radial profile of axion field $\phi$ for case
$\alpha=0.25$ at $\epsilon_{0}=650$ (upper panel), $800$ (middle
panel) and $1000$ (down panel) Mev/fm$^3$ correspondingly.}
\end{figure}

\begin{figure}
\includegraphics[scale=0.32]{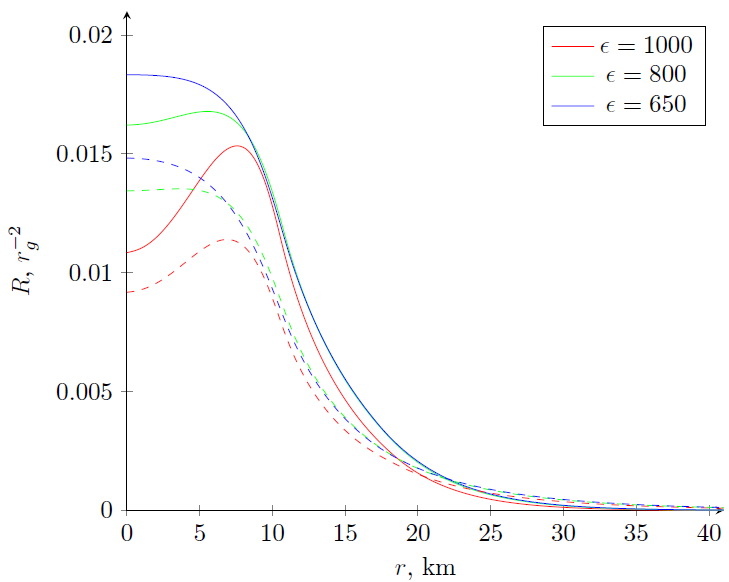}\\
\includegraphics[scale=0.32]{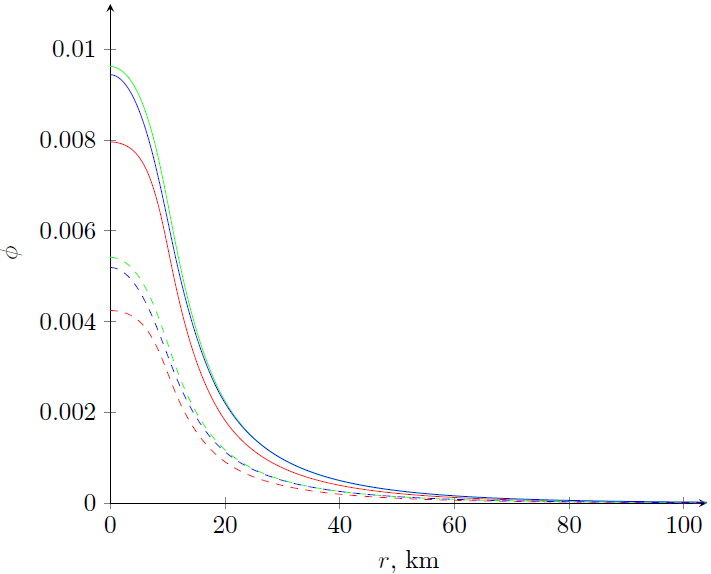}\\
\includegraphics[scale=0.32]{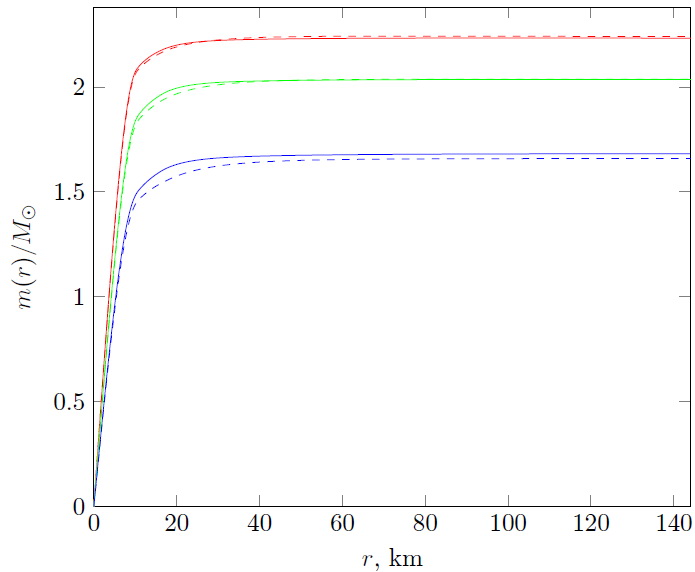}
\caption{Scalar curvature, axion field and mass profiles for some
densities in the center of star in a case of $\beta=250$ for two
values of $\alpha$: $0.25$ (solid lines) and $2.5$ (dashed
lines).}
\end{figure}

\begin{figure}
\includegraphics[scale=0.32]{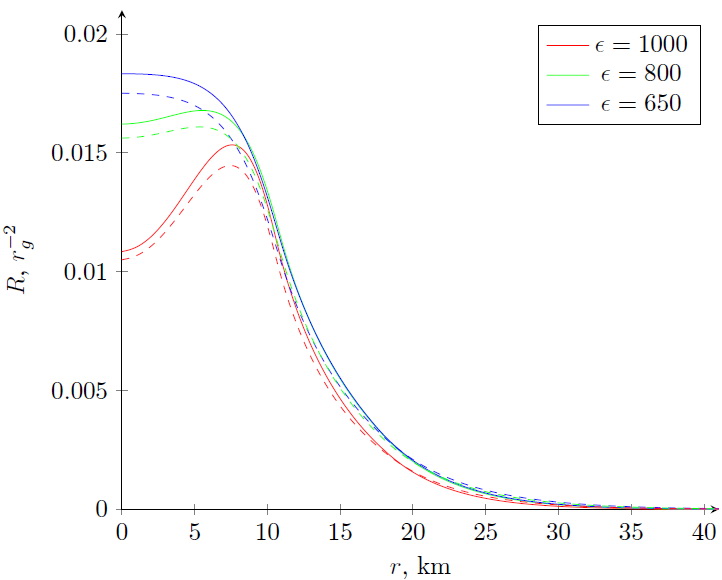}\\
\includegraphics[scale=0.32]{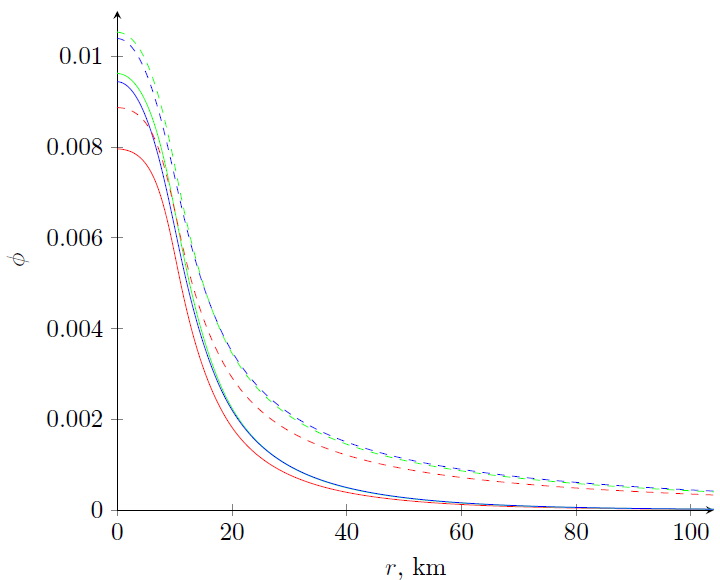}\\
\includegraphics[scale=0.32]{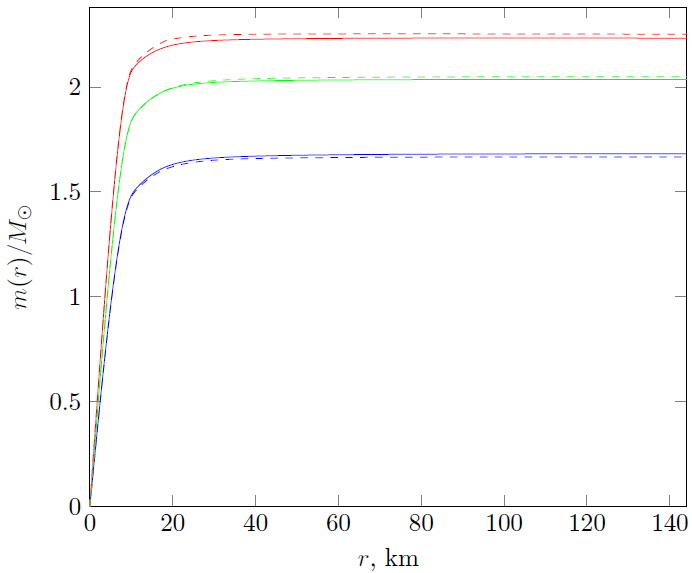}
\caption{Scalar curvature, axion field and mass profiles for some
densities in the center of star in a case of $\beta=250$ for two
values of $m_a$: $0.1$ (solid lines) and $0.01$ (dashed lines).}
\end{figure}

\begin{table}
\begin{tabular}{|c|c|c|c|c|c|}
\hline $\epsilon_{0}$, MeV/fm$^{3}$  & $\beta$, $r_{g}^{2}$   &
$M/M_{\odot}$, &
$R_{s}$, km      & $R_{0}$, $r_{g}^{-2}$    & $\phi_{0}$ \\
\hline
         & GR & 1.64  & 11.25 & 0.0674 & -     \\
         & 0 & 1.68  & 11.40 & 0.0674 & -     \\
   650   & 100 & 1.65  & 11.38 & 0.0273 & 0.0097     \\
         & 250 & 1.67  & 11.42 & 0.0183 & 0.0094     \\
         & 1000& 1.73  & 11.48 & 0.0077 & 0.0076     \\
         \hline
         & GR & 1.92  & 11.03 & 0.0146 & -     \\
         & 0 & 1.98  & 11.15 & 0.0137 & -     \\
   800   & 100 & 2.00  & 11.26 & 0.0238 & 0.0097     \\
         & 250 & 2.04  & 11.33 & 0.0162 & 0.0096     \\
         & 1000 & 2.12  & 11.44 & 0.0071 & 0.0077    \\
         \hline
         & GR & 2.10  & 10.68 & -0.0889 & -     \\
         & 0 & 2.14  & 10.80 & -0.0899 & -     \\
   1000  & 100 & 2.19  & 10.89 & 0.0142 & 0.0077     \\
         & 250 & 2.22  & 10.95 & 0.0108 & 0.0080     \\
         &1000 & 2.30  & 11.07 & 0.0051 & 0.0068     \\
\hline
\end{tabular}
\caption{Parameters of compact stars (mass and radius)  in General
Relativity, pure $R^2$ gravity ($\alpha=0.25$) and for several
values of $\beta$ ($\alpha=0.25$) in the model with axion field
for three values of energy density in the center $\epsilon_{0}$.
The corresponding values of curvature and axion field in the
center of star are given also. For massive stars one can see that
increase of mass for $\beta=10^3$ consists of $\sim 0.2M_{\odot}$
(in comparison with General Relativity) for the same density in
the center of star.} \label{Table1}
\end{table}

\section{Results}

We considered in detail the model with the action
\begin{equation}\label{our}
\mathcal{S}=\int d^{4}x\sqrt{-g}\left(\frac{R+\alpha
R^2}{16\pi}+\beta\frac{R^2\phi}{16\pi}-\frac{1}{2}\partial_{\mu}\phi\partial^{\mu}\phi
-V(\phi)\right).
\end{equation}

Following to \citealp{Marsh} we consider only small deviations from
the potential minimum. Therefore, the leading term is
$$V(\phi)\simeq\frac{1}{2}{m^{2}_{a}}\phi^{2}.$$

The behavior of axion scalar  is mainly governed by
$m_{a}^{2}\phi^{2}$ term. Another key moment of this model is the
influence of scalar field on the dependence of scalar curvature
from radial coordinate.

For axion mass we take the  value corresponding to Compton
wavelength $10r_{g}$ where $r_g$ is gravitational radius of Sun
($2.95$ km). For the parameter $\beta$ the range $100<\beta<1000$
in units of $r_{g}^2$ is explored.

First of all, let us consider the case when the parameter $\alpha$
is relatively small (for example $\alpha=0.25$). The mass-radius
diagram can be seen on Fig. 1 with the dependence of stellar mass
from the energy density in the center of star. For massive stars
one can see (Table I) that increase of mass for $\beta=10^3$
consists of $\sim 0.2M_{\odot}$ or $\sim 10$\% (in comparison with
General Relativity) for the same density in the center of star.
The radius of star also increases and therefore for given radius
the increase of mass looks even bigger. Therefore one can expect
that in such modified gravity the fraction of supermassive neutron
stars (with $M>2.0M_{\odot}$) should increase. Another interesting
point is the maximal possible mass for given equation of state.
For APR equation we have that $M_{max}=$2.31$M_{\odot}$ at
$\beta=10^3$ in comparison with $M_{max}=2.16M_{\odot}$) in
General Relativity. Note that mass increases with $\beta$
nonlinearly and for very large  $\beta$ this increase does not
exceed significantly the result obtained for $\beta=1000$. {In
General Relativity one can obtain compact stars with $M\sim 2.3
M_{\odot}$ and $R_s\sim 11$ km only for very stiff equations of
state usually treated as unrealistic.}

It is interesting to consider the mass profile $m(r)$ for various
parameters of model. On Fig. 2 the $m(r)$ for three values of
$\epsilon_{0}$ (650, 800 and 1000 MeV/fm$^3$) is depicted. The
increase of gravitational mass of star for distant observer occurs
due to non-trivial behavior of scalar curvature outside the star
(Fig. 3). For GR $R=0$ outside the star and for simple $R^2$-model
of gravity scalar curvature decreases quickly outside the star.
Due to axion field (see Fig. 4) the damping of scalar curvature
became smoother.

The increase of parameter $\alpha$ in considered area
($0.25<\alpha<2.5$) does not lead to significant consequences for
stellar masses. For example on Fig. 5 the radial profiles of
scalar curvature, axion field and mass are depicted for fixed
value of $\beta=250$ and two values of $\alpha$: $0.25$ and $2.5$.
One can see that increase of $\alpha$  leads to some decrease of
axion field and curvature (therefore contribution of term $\beta
R^2\phi$ decreases). {Note that we don't consider the case of very
large $\alpha$ ($\sim O(100)$) by the following reason. One can
see that value of curvature within star is $\sim 0.01-0.015$.
Therefore term $\alpha R^2$ for $\alpha\sim O(100)$ will be larger
in comparison with $R$. Therefore from physical viewpoint such
values of $\alpha$ seems unrealistic.}

One can also mention about the influence of axion mass on the
solution of gravitational equations. We considered  $m_{a}=0.01$
(corresponding to Compton wavelength $100r_{g}$ and found that in
this case only solution of scalar field outside the star changes
considerably (see Fig. 6) but scalar curvature for $r>50$ km is
very close to zero and therefore contribution of term $R\phi^2$ is
negligible.

{We also considered in our preliminary calculations the potential
with $\sim \phi^4$ term but in this case deviations from simple
model are very negligible.}

\section{Conclusion}

We investigated  neutron stars in axion $R^2$ gravity with
non-minimal curvature-axion coupling in form $\sim R^2\phi$. Our
main result is possible increase of stellar mass due to axion
presence. We obtained the increase of mass $\sim 0.2M_{\odot}$ for
massive stars in the case of $\beta=1000$. This value is
sufficient for possible observational indication of such model.
{The
  star radius increases not so considerably ($\sim 400$ m for
$\beta=1000$). Therefore, axion $F(R)$ gravity under consideration
may explain the possible existence of supermassive neutron stars
($M>2.2M_\odot$) compact as in general relativity at the same
time. There are some indications in favor of the existence of such
neutron stars (for example the possible masses of B1957+20
(\citealp{Kerk}) and 4U 1700-377 (\citealp{Clark}) are $M\sim 2.4
M_{\odot}$).}

Axion field affects on behavior of scalar curvature inside and
outside star in comparison with General Relativity and vacuum
$R^2$ gravity. Increase of mass for distant observer occurs due to
``gravitational sphere'' outside the star with nonzero curvature.
The  star radius increases and mass confined inside the stellar
surface decreases in comparison with General Relativity but the
contribution of gravitational sphere overcompensates this
decrease. In contrast to simple square gravity increase of
gravitational mass is relatively equal for various values of
density in the center of star up to the masses close to maximum.
Therefore, the increase of maximal mass takes place. For instance,
we obtained that in the case of APR equation of state maximal mass
is 2.31$M_{\odot}$ instead of 2.15$M_{\odot}$ in General
Relativity for some choice of parameters. Increase of radius also
take place but it is not so significant and hardly observable.

Our calculations show also interesting effect of some
``compensation'' between two terms, i.e. $\alpha R^2$ and $\beta
\phi R^2$. If $\alpha$ increases the contribution of second term
decreases due to damping mean value of curvature and axion field.
As the consequence we have no possibility to discriminate between
various solutions corresponding to various parameters.

Characteristic scale of curvature damping in pure $R^2$ gravity is
$\sim \alpha^{1/2}$. We considered the case when this value is
smaller in comparison with Compton wavelength of axion field ($10
r_{g}$). For very large Compton wavelength ($\sim 100 r_{g}$, for
example) we have no observable consequences on masses and radii of
stellar configurations. Only radius of axion ``galo'' around the
star grows.

Although, for illustration we used the star models based on
well-known APR equation of state our calculations  lead to
qualitatively similar results for other realistic choices of
  equation of state for dense matter in neutron stars.
Finally, one can expect that combined effect of axion dark matter
and modified gravity maybe quite significant in stellar
astrophysics at strong gravitational regime.

{One should mention recent paper (\citealp{Riley}) in which
authors found promising mass-radius posteriors for mass and radius
of pulsar PSR J0030+0451. We think that these results on current
stage can exclude some equations of state for which radius of star
with mass $M\sim 1.3-1.5M_\odot$ differs significantly from narrow
range. In the light of these data APR equation of state is under
question in General Relativity and as consequence in our model
because for $M=1.3-1.5M_{\odot}$ possible value of radius differs
from GR value negligibly (this difference is $\sim 100$ m for
model with axions and this value is less in comparison with error
of measurements ($\sim$1 km)). For equation of state describing
these data the picture is the same. In this case if GR fits these
data well then our theory does the same.}

{One can ask how would one differentiate between considered model
of simple $f(R)$ gravity with axions and simply for example  the
case with different equation of state for dense matter?
Eventually, the observational indication towards our model may
soon appear if the expected experiments may confirm that axion is
indeed the dark matter. From another side, R2 gravity gives the
best realistic candidate for inflation. Again, if future more
precise Planck/BICEP data will confirm axion R2 inflation that
will be the best proof of viability of current model for
supermassive neutron stars.}

Acknoweledgments. This work is partly supported by  MINECO
(Spain), FIS2016-76363-P, by COST Action PHAROS (CA16214), by
project 2017 SGR247 (AGAUR, Catalonia) (SDO). AVA thanks the
Program 5-100 (IKBFU, Russia).

\bibliographystyle{mnras}

\begin{thebibliography}{}

\bibitem[\protect\citeauthoryear{CDMS II Collaboration, et al.}{2010}]{Ahmed} CDMS II Collaboration, et al., 2010, Sci, 327, 1619

\bibitem[\protect\citeauthoryear{Alavirad \& Weller}{2013}]{Alavirad2013} Alavirad H., Weller J.~M., 2013, PhRvD, 88, 124034

\bibitem[\protect\citeauthoryear{Alcubierre}{2008}]{Alcubierre} Alcubierre M., 2008, itnr.book

\bibitem[\protect\citeauthoryear{Arapo{\v{g}}lu, Deliduman \& Ek{s}i}{2011}]{Arapoglu2011} Arapo{\v{g}}lu S., Deliduman C., Ek{s}i K.~Y., 2011,
JCAP, 2011, 020

\bibitem[\protect\citeauthoryear{Astashenok, Capozziello \& Odintsov}{2013}]{Astashenok2013} Astashenok A.~V., Capozziello S., Odintsov S.~D., 2013, JCAP, 2013, 040

\bibitem[\protect\citeauthoryear{Astashenok, Capozziello \& Odintsov}{2014}]{Astashenok2014} Astashenok A.~V., Capozziello S., Odintsov S.~D., 2014, PhRvD, 89, 103509

\bibitem[\protect\citeauthoryear{Astashenok, Capozziello \& Odintsov}{2015}]{Astashenok2015} Astashenok A.~V., Capozziello S., Odintsov S.~D., 2015, Ap\&SS, 355, 333

\bibitem[\protect\citeauthoryear{Astashenok, Odintsov \& de la Cruz-Dombriz}{2017}]{Astashenok2017} Astashenok A.~V., Odintsov S.~D., de la Cruz-Dombriz {\'A}., 2017, CQGra, 34, 205008

\bibitem[\protect\citeauthoryear{Avignone, Creswick \& Vergados}{2018}]{Avignone} Avignone F.~T., Creswick R.~J., Vergados J.~D., 2018, arXiv, arXiv:1801.02072

\bibitem[\protect\citeauthoryear{Balakin \& Ni}{2010}]{AX-3} Balakin A.~B., Ni W.-T., 2010, CQGra, 27, 055003

\bibitem[\protect\citeauthoryear{Balakin, Bochkarev \& Tarasova}{2012}]{AX-4} Balakin A.~B., Bochkarev V.~V., Tarasova N.~O., 2012, EPJC, 72, 1895

\bibitem[\protect\citeauthoryear{Balakin, Muharlyamov \& Zayats}{2014}]{AX-5} Balakin A.~B., Muharlyamov R.~K., Zayats A.~E., 2014, EPJD, 68,
159

\bibitem[\protect\citeauthoryear{Baumgarte \& Shapiro}{2010}]{Shapiro} Baumgarte T.~W., Shapiro S.~L., 2010, nure.book

\bibitem[\protect\citeauthoryear{Brada{\v{c}}, et al.}{2008}]{Bradac} Brada{\v{c}} M., et al., 2008, ApJ, 687, 959

\bibitem[\protect\citeauthoryear{Capozziello \& Fang}{2002}]{Capozziello1} Capozziello S., Fang L.~Z., 2002, IJMPD, 11, 483

\bibitem[\protect\citeauthoryear{Capozziello \& de Laurentis}{2011}]{Capoz} Capozziello S., de Laurentis M., 2011, PhR, 509, 167

\bibitem[\protect\citeauthoryear{Capozziello, et al.}{2011}]{Capozz2011} Capozziello S., de Laurentis M., Odintsov S.~D., Stabile A., 2011, PhRvD, 83, 064004

\bibitem[\protect\citeauthoryear{Capozziello, et al.}{2012}]{Capozz2012} Capozziello S., de Laurentis M., de Martino I., Formisano M., Odintsov S.~D., 2012, PhRvD, 85, 044022

\bibitem[\protect\citeauthoryear{Capozziello, et al.}{2016}]{Capozz-1} Capozziello S., De Laurentis M., Farinelli R., Odintsov S.~D., 2016, PhRvD, 93, 023501

\bibitem[\protect\citeauthoryear{Caputo, et al.}{2019}]{Caputo-2} Caputo A., Regis M., Taoso M., Witte S.~J., 2019, JCAP, 2019, 027

\bibitem[\protect\citeauthoryear{Caputo, Garay \& Witte}{2018}]{Caputo-3} Caputo A., Garay C.~P., Witte S.~J., 2018, PhRvD, 98, 083024

\bibitem[\protect\citeauthoryear{Caputo}{2019}]{Caputo} Caputo A., 2019, PhLB, 797, 134824

\bibitem[\protect\citeauthoryear{Carroll, et al.}{2004}]{Turner} Carroll S.~M., Duvvuri V., Trodden M., Turner M.~S., 2004, PhRvD, 70, 043528

\bibitem[\protect\citeauthoryear{Cheoun, et al.}{2013}]{Cheoun2013} Cheoun M.-K., Deliduman C., G{\"u}ng{\"o}r C., Kele{s} V., Ryu C.~Y., Kajino T., Mathews G.~J., 2013, JCAP, 2013, 021

\bibitem[\protect\citeauthoryear{Cicoli, Guidetti \& Pedro}{2019}]{Cicoli} Cicoli M., Guidetti V., Pedro F.~G., 2019, JCAP, 2019, 046

\bibitem[\protect\citeauthoryear{Clark, et al.}{2002}]{Clark} Clark J.~S., Goodwin S.~P., Crowther P.~A., Kaper L., Fairbairn M., Langer N., Brocksopp C., 2002, A\&A, 392, 909

\bibitem[\protect\citeauthoryear{Clowe, et al.}{2006}]{Clowe} Clowe D., Brada{\v{c}} M., Gonzalez A.~H., Markevitch M., Randall S.~W., Jones C., Zaritsky D., 2006, ApJL, 648, L109

\bibitem[\protect\citeauthoryear{de la Cruz-Dombriz \& S{\'a}ez-G{\'o}mez}{2012}]{Cruz} de la Cruz-Dombriz A., S{\'a}ez-G{\'o}mez D., 2012, Entrp, 14, 1717

\bibitem[\protect\citeauthoryear{Davis, McCabe \& B{\oe}hm}{2014}]{Davis} Davis J.~H., McCabe C., B{\oe}hm C., 2014, JCAP, 2014, 014

\bibitem[\protect\citeauthoryear{Davis}{2015}]{Davis2} Davis J.~H., 2015, IJMPA, 30, 1530038

\bibitem[\protect\citeauthoryear{Anisimov \& Dine}{2005}]{AX-2} Anisimov A., Dine M., 2005, JCAP, 2005, 009

\bibitem[\protect\citeauthoryear{Du, et al.}{2018}]{ADMX} Du N., et al., 2018, PhRvL, 120, 151301

\bibitem[\protect\citeauthoryear{Farinelli, et al.}{2014}]{Capozz-2} Farinelli R., De Laurentis M., Capozziello S., Odintsov S. D., 2014, MNRAS, 440, 2909

\bibitem[\protect\citeauthoryear{Feola, et al.}{2019}]{Capozz-3} Feola P., Jimenez Forteza X., Capozziello S., Cianci R., Vignolo S., 2019, arXiv, arXiv:1909.08847

\bibitem[\protect\citeauthoryear{Fukunaga, Kitajima \& Urakawa}{2019}]{Fukunaga} Fukunaga H., Kitajima N., Urakawa Y., 2019, JCAP, 2019, 055

\bibitem[\protect\citeauthoryear{Gourgoulhon}{2007}]{Gourg-2} Gourgoulhon E., 2007, arXiv, gr-qc/0703035

\bibitem[\protect\citeauthoryear{Gourgoulhon}{2010}]{Gourg} Gourgoulhon E., 2010, arXiv, arXiv:1003.5015

\bibitem[\protect\citeauthoryear{Henning, et al.}{2018}]{ABRACADABRA} [ABRACADABRA
Collaboration], 2018, DOI:
10.3204/DESY-PROC-2017-02/henning\_reyco

\bibitem[\protect\citeauthoryear{van Kerkwijk, Breton \& Kulkarni}{2011}]{Kerk} van Kerkwijk M.~H., Breton R.~P., Kulkarni S.~R., 2011, ApJ, 728, 95

\bibitem[\protect\citeauthoryear{Khlopov, Sakharov \& Sokoloff}{1999}]{Sakharov-3} Khlopov M.~Y., Sakharov A.~S., Sokoloff D.~D., 1999, NuPhS, 72, 105

\bibitem[\protect\citeauthoryear{Lawson, et al.}{2019}]{AX-1} Lawson M., Millar A.~J., Pancaldi M., Vitagliano E., Wilczek F., 2019, PhRvL, 123, 141802

\bibitem[\protect\citeauthoryear{Markevitch, et al.}{2003}]{Markevitch} Markevitch M., et al., 2003, ApJ, 583, 70

\bibitem[\protect\citeauthoryear{Marsh}{2016}]{Marsh} Marsh D.~J.~E., 2016, PhR, 643, 1

\bibitem[\protect\citeauthoryear{Marsh, et al.}{2017}]{Marsh-2} Marsh M.~C.~D., Russell H.~R., Fabian A.~C., McNamara B.~R., Nulsen P., Reynolds C.~S., 2017, JCAP, 2017, 036

\bibitem[\protect\citeauthoryear{Nojiri \& Odintsov}{2003}]{Odintsov1} Nojiri S., Odintsov S.~D., 2003, PhRvD, 68, 123512

\bibitem[\protect\citeauthoryear{Nojiri \& Odintsov}{2003}]{Nojiri} Nojiri S., Odintsov S.~D., 2003, PhLB, 576, 5

\bibitem[\protect\citeauthoryear{Nojiri, Odintsov \& Oikonomou}{2017}]{Nojiri-4} Nojiri S., Odintsov S.~D., Oikonomou V.~K., 2017, PhR, 692, 1

\bibitem[\protect\citeauthoryear{Nojiri \& Odintsov}{2011}]{Nojiri-5} Nojiri S., Odintsov S.~D., 2011, PhR, 505, 59

\bibitem[\protect\citeauthoryear{Odintsov \& Oikonomou}{2019}]{Oikonomou-19} Odintsov S.~D., Oikonomou V.~K., 2019, PhRvD, 99, 104070

\bibitem[\protect\citeauthoryear{Odintsov \& Oikonomou}{2019}]{Oikonomou} Odintsov S.~D., Oikonomou V.~K., 2019, PhRvD, 99, 104070

\bibitem[\protect\citeauthoryear{Olmo}{2011}]{Olmo-2} Olmo G.~J., 2011, IJMPD, 20, 413

\bibitem[\protect\citeauthoryear{Olmo, Rubiera-Garcia \& Wojnar}{2019}]{Olmo} Olmo G.~J., Rubiera-Garcia D., Wojnar A., 2019, arXiv, arXiv:1912.05202

\bibitem[\protect\citeauthoryear{Perlmutter, et al.}{1999}]{Perlmutter} Perlmutter S., et al., 1999, ApJ, 517, 565

\bibitem[\protect\citeauthoryear{Ouellet, et al.}{2019}]{Quellet} Ouellet J.~L., et al., 2019, PhRvL, 122, 121802

\bibitem[\protect\citeauthoryear{Rezzolla, et al.}{2018}]{Rezzolla} Rezzolla L., Pizzochero P., Jones D.~I., Rea N., Vida{\~n}a I., 2018, ASSL..457

\bibitem[\protect\citeauthoryear{Riess, et al.}{1998}]{Riess-1} Riess A.~G., et al., 1998, AJ, 116, 1009

\bibitem[\protect\citeauthoryear{Riess, et al.}{2004}]{Riess-2} Riess A.~G., et al., 2004, ApJ, 607, 665

\bibitem[\protect\citeauthoryear{Riley, et al.}{2019}]{Riley} Riley T.~E., et al., 2019, ApJL, 887, L21

\bibitem[\protect\citeauthoryear{Robertson, Massey \& Eke}{2017}]{Robertson} Robertson A., Massey R., Eke V., 2017, MNRAS, 465, 569

\bibitem[\protect\citeauthoryear{Roszkowski, Sessolo \& Trojanowski}{2018}]{Roszkowski} Roszkowski L., Sessolo E.~M., Trojanowski S., 2018, RPPh, 81, 066201

\bibitem[\protect\citeauthoryear{Rozner, et al.}{2019}]{Rozner} Rozner M., Grishin E., Ginat Y.~B., Igoshev A.~P., Desjacques V., 2019, arXiv, arXiv:1904.01958

\bibitem[\protect\citeauthoryear{Safdi, Sun \& Chen}{2019}]{Safdi} Safdi B.~R., Sun Z., Chen A.~Y., 2019, PhRvD, 99, 123021

\bibitem[\protect\citeauthoryear{Sakharov \& Khlopov}{1994}]{Sakharov} Sakharov A.~S., Khlopov M.~Y., 1994, PAN, 57, 651

\bibitem[\protect\citeauthoryear{Sakharov, Sokoloff \& Khlopov}{1996}]{Sakharov-2} Sakharov A.~S., Sokoloff D.~D., Khlopov M.~Y., 1996, PAN, 59, 1005

\bibitem[\protect\citeauthoryear{Schumann}{2019}]{Schumann} Schumann M., 2019, JPhG, 46, 103003



\end{thebibliography}

\label{lastpage}
\end{document}